\documentclass[twocolumn,prl,floatfix,superscriptaddress]{revtex4-1}
\usepackage[labelfont=bf,justification=raggedright,singlelinecheck=false]{caption}
\usepackage{wrapfig, graphicx, subfig}
\usepackage[nottoc, numbib]{tocbibind}
\usepackage[utf8]{inputenc}
\usepackage{physics}
\usepackage{amssymb, amsthm}
\usepackage{siunitx}
\usepackage{nicefrac}
\usepackage{hyperref}
\hypersetup{
    colorlinks = false,
    pdfborder = {0 0 0 [3 3]}
}
\usepackage{tikz}
\begin{document}

\title{Revival and expansion of the theory of coherent lattices}

\author{Dmitry Kouznetsov}
\thanks{Corresponding author \\ dmitry.kouznetsov@imec.be}
\affiliation{KU Leuven, Dept. of Physics and Astronomy, Research unit Quantum Solid-State Physics, B-3001 Leuven, Belgium}
\affiliation{imec, Kapeldreef 75, B-3001 Leuven, Belgium}
\author{Qingzhong Deng}
\affiliation{KU Leuven, Dept. of Physics and Astronomy, Research unit Quantum Solid-State Physics, B-3001 Leuven, Belgium}
\affiliation{imec, Kapeldreef 75, B-3001 Leuven, Belgium}
\author{Pol Van Dorpe}
\affiliation{imec, Kapeldreef 75, B-3001 Leuven, Belgium}
\affiliation{KU Leuven, Dept. of Physics and Astronomy, Research unit Quantum Solid-State Physics, B-3001 Leuven, Belgium}
\author{Niels Verellen}
\thanks{Corresponding author \\ niels.verellen@imec.be}
\affiliation{imec, Kapeldreef 75, B-3001 Leuven, Belgium}

\date{\today}

\begin{abstract}
    An effective way to design structured coherent wave interference patterns that builds on the theory of coherent lattices, is presented.
    The technique combines prime number factorization in the complex plane with moiré theory to provide a robust way to design structured patterns with variable spacing of intensity maxima.
    In addition, the proposed theoretical framework facilitates an elegant computation of previously unexplored high-order superlattices both for the periodic and quasiperiodic case.
    A number of beam configurations highlighting prime examples of patterns for lattices with three-, four-, and fivefold symmetry are verified in a multibeam interference experiment.
\end{abstract}

\maketitle

Structured interference patterns of coherent waves, known as \emph{coherent lattices}, play a key role in a wide range of applications.
In particular, coherent lattices are prominent in structured illumination microscopy~\cite{Gustafsson_2000,York_2012}, fabrication of microstructures such as photonic~\cite{Campbell_2000} or plasmonic crystals~\cite{Kasture_2014}, optical trapping in the life sciences~\cite{Ashkin_1986} and quantum research~\cite{Greiner_2002,Mace_2016}.
In all these applications, the periodicity of the patterns has been of the same order as the wavelength $\lambda$.
The intensity maxima defining the periodicity are thus too closely spaced to be used as individually resolvable excitation foci or trapping potentials.
Sparse multifocal excitation fields have been proposed more than fifteen years ago by Betzig~\cite{Betzig_2005} but have gone virtually unnoticed outside the microscopy community, where the implementation of coherent lattices in light-sheet microscopy was limited to a small subset of configurations~\cite{Chen_2014}.
More recent advancements in microscopy are targeted toward lens-free implementations that use photonic integrated circuit (PIC) technology.
Also here, coherent lattices can prove to be pivotal as it allows for a planar configuration of input waves to generate sparse optical superlattices.

The applicability of structured interference patterns spans beyond photonics.
In materials science research, for example, acoustic interference patterns enable measuring piezoelectric transduction efficiency~\cite{Zheng_2018}.
Furthermore, a subset of sparse lattices was discovered in thermal convection~\cite{Rogers_2000},
and Faraday wave experiments~\cite{Porter_2004,Rucklidge_2009}, but the connection to the theory of coherent lattices was not yet made to our knowledge.

Sparse and composite coherent lattices are designed by cycling through combinatorial lattice symmetry configurations, as described in Ref.~\cite{Betzig_2005}.
By observing that both sparse and composite lattices are rotational moiré superlattices~\cite{Amidror_2009}, we present a new model that relies on symmetries in number fields.
The proposed model successfully predicts coherent superlattices ranked by order of moiré transformation and is equally valid for quasiperiodic lattices, which were previously not treated~\cite{Betzig_2005}.
The rotational moiré pattern synthesis consists of superposing periodic lattices with a relative rotation, further referred to as \emph{twist}.
It has been demonstrated that the interference patterns obtained by concentrically arranging coherent wave sources at equal distances around a circle have (quasi-)periodic symmetry~\cite{Senechal_2009}.
Starting from these simple interference patterns, the moiré superpattern is obtained by averaging the wavefront over a finite group of transformations.
We will show that the values of these \emph{twist} angles are encoded in the complex prime factorization.
Furthermore, we present experimental results validating this \emph{integer lattice method} approach.

\begin{figure}[!ht]
    \centering
    \includegraphics[width=\linewidth,keepaspectratio]{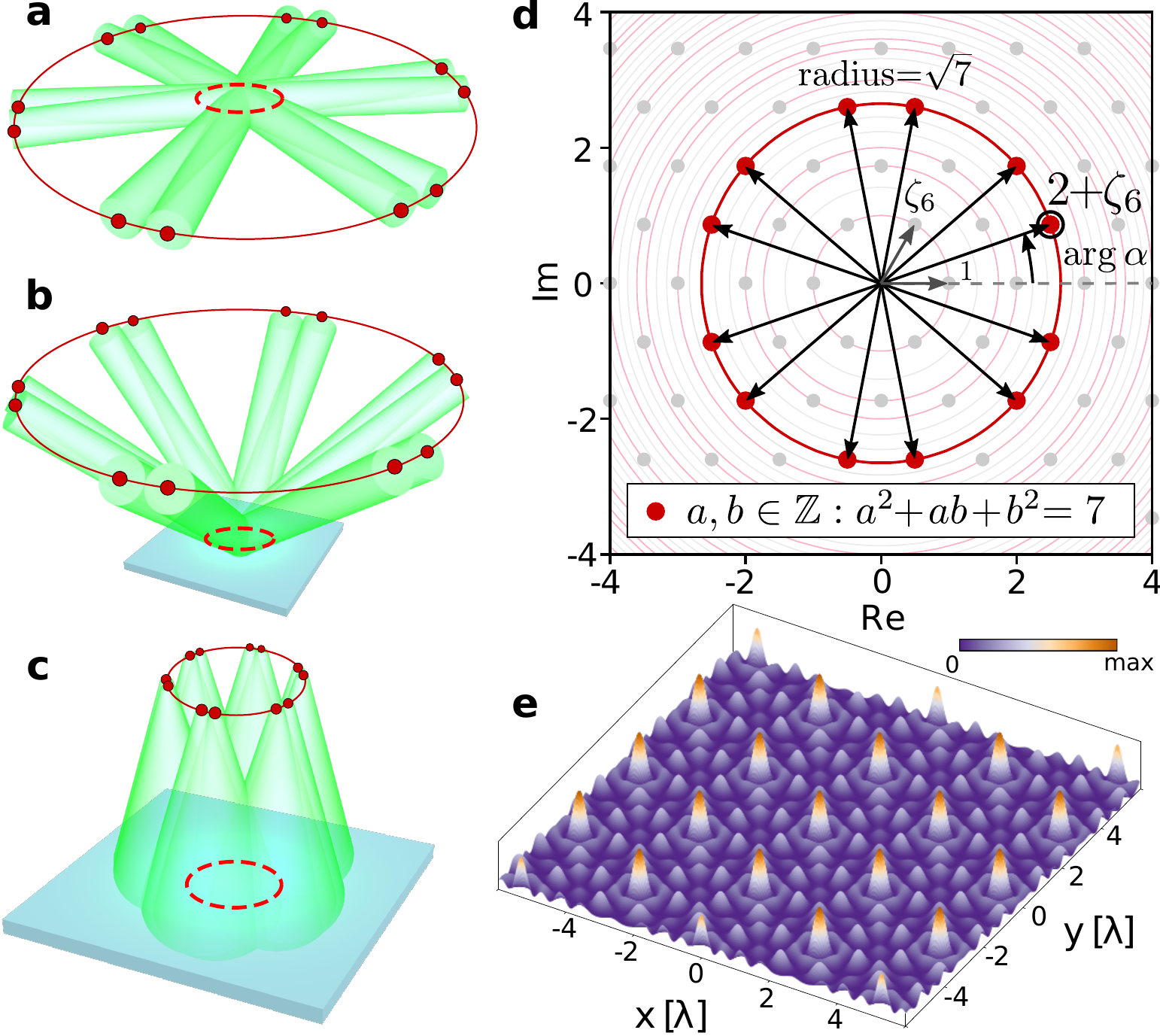}
    \caption{Laser beam configurations for generation of optical coherent lattices.
    The configuration can be planar (a), tilted with respect to the horizontal (b), or using diffraction cones (c).
    The orientations of the beams are determined from $\alpha=a+b\zeta_6\in\mathbb{Z}[\zeta_6]$ with $\zeta_6=e^{\nicefrac{2\pi i}{6}}$ and field norm $n=7$, highlighted in red (d).
    In the region where all beams overlap a triangular coherent lattice will form (e).}
  \label{fig:mbi}
\end{figure}
\emph{Light as scalar waves} ---
Arbitrary coherent waves such as acoustic, two-layer interfacial waves, and monoenergetic matter waves, can be described by plane wave functions.
This also holds true for electromagnetic waves when disregarding polarization.
We define a plane wave as a function $f:\mathbb{C}\rightarrow\mathbb{C}$ propagating along the $y$-axis with the wavefront parallel to the $x$-axis
\begin{equation}
    \label{eq:plane_wave}f(z) = e^{2\pi i \Im(z)}\,,
\end{equation}
where $z=x+i y \in \mathbb{C}$.
The plane wave is expressed in units of wavelength $\lambda$.
If $G$ is a finite set of transformations of the complex plane with $N$ elements, the average of Eq. \eqref{eq:plane_wave} over the elements of $G$ will be defined as
\begin{equation}
    \label{eq:averaged_plane_waves}\hat{f}(z) = \frac{1}{N} \sum _{g\in G} f(g\cdot z) \,.
\end{equation}
This equation corresponds to the normalized total superposition of the plane waves oriented according to the elements of $G$.

\begin{figure*}[!ht]
    \centering
    \includegraphics[width=\textwidth,keepaspectratio]{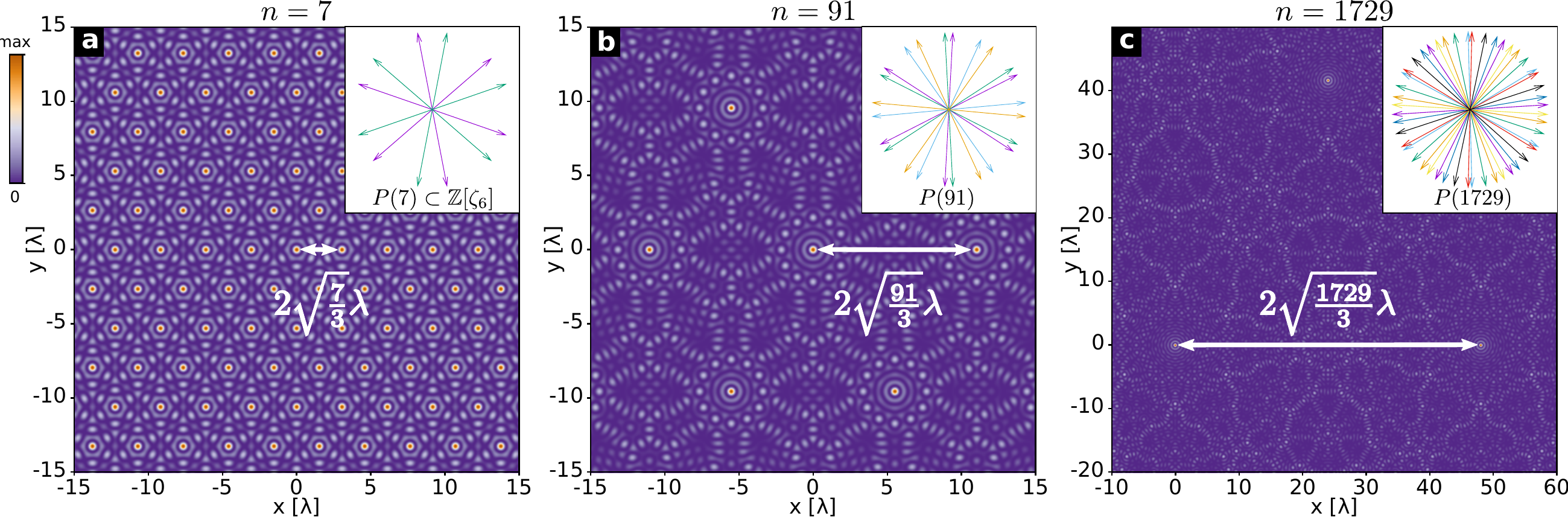}
    \caption{Concyclic integer lattice points in $\mathbb{Z}[\zeta_6]$ (represented by arrows and grouped by color according to multiples of the units) with the generated coherent lattices showing increasing periodicity.
    The corresponding interference patterns are calculated for (a) $n=7$, (b) $n=91$ and (c) $n=1729$ with respectively $12, 24$ and $48$ wave components.}
    \label{fig:size}
\end{figure*}
Specific to electromagnetic waves, the intensity of the interference pattern is obtained from the complex amplitude (Eq. \eqref{eq:averaged_plane_waves}) by multiplying it with its complex conjugate~\cite{Fowles_1989}:
\begin{equation}
    \label{eq:intensity}I(z) = \hat{f}(z) \hat{f}^*(z) \,.
\end{equation}
Notably, the choice of $G$ will determine the symmetry of the resulting pattern.

\emph{Plane wave orientations from integer lattices} ---
The common way to obtain (quasi-)periodic patterns is to orient wavefronts at equidistant angles.
In an optical setup this can be realized for various light beam configurations ranging from planar (Fig.~\ref{fig:mbi}a) such as in systems with 2D confined wave propagation like PICs, to tilted (Fig.~\ref{fig:mbi}b) such as in conventional interference setups, to diffraction cones (Fig.~\ref{fig:mbi}c) such as in photolithography using photomasks.
Configuring beams in an equally spaced circular configuration corresponds to choosing $G$ to be a cyclic group.
The generator of the cyclic group $\zeta_m=e^{\nicefrac{2\pi i}{m}}$, where $m\geq3$ is an integer,  generates an algebraic field $\mathbb{Q}(\zeta_m)$ for which regular structures such as integer lattices can be defined~\cite{Washington_1997}.
We will make the key observation that finding the plane wave orientations from these integer lattices is equivalent to selecting wave vectors as was established in Ref.~\cite{Betzig_2005}.

We focus on integer lattices in cyclotomic fields $\mathbb{Q}(\zeta_m)\supset\mathbb{Z}[\zeta_m]=\{a+b\,\zeta_m\mid a,b\in\mathbb{Z}\}$, where $m$ is an integer \cite{Washington_1997}.
In the complex plane the only 2D periodic lattices that are allowed by the crystallographic restriction theorem are the rectangular and triangular lattices \cite{Senechal_2009}.
Our argument can be further restricted to the case where $m$ is even because \mbox{$\mathbb{Q}(\zeta_m)=\mathbb{Q}(\zeta_{2m})$} for odd $m$, as is detailed in Re.~\cite{Washington_1997}.

For these specific cases, the representations of an integer $n$ are known to be concyclic.
Each lattice point on a circle with radius $\sqrt{n}$ corresponds to a representation of $n$ as a product of $a+\zeta_m b$ and its conjugate.
To find all concyclic lattice points for a given $n$, we compute the field norm, for any $\alpha\in\mathbb{Z}[\zeta_m]$ defined \cite{Washington_1997} as
\begin{equation}
    N(\alpha) = \alpha \bar\alpha = \begin{cases} a^2 + b^2 & \text{for } m=4 \\ a^2 + ab + b^2 & \text{for } m=6 \,.\\ \end{cases}
\end{equation}
Since $\mathbb{Z}[\zeta_m]$ with $m=4$ or $6$ is Euclidean and hence a unique factorization domain, the number of points on each circle with radius $\sqrt{n}$ is equal to the number of combinations of the complex factors of the prime factorization of $n$ in the considered number ring \cite{Hardy_2008}.

The set of points
\begin{equation}
    P(n) = \{ N(\alpha)=n \mid \alpha \in \mathbb{Z}[\zeta_m] \}\,, \label{eq:concyclic}
\end{equation}
normalized to have unit magnitude, are used as input for calculating the intensity in Eq.~\eqref{eq:intensity}, i.e. setting $G=n^{-1/2}P(n)$.
As an example, in Fig.~\ref{fig:mbi}d, the triangular integer lattice is shown with points selected by setting \mbox{$n=7$}.
The obtained corresponding coherent lattice is shown in Fig.~\ref{fig:mbi}e.
Note that the normalization step ensures that all chosen integer lattice points have unit magnitude and act as pure rotations, fulfilling the coherence criterion.

\emph{Relation with moiré} ---
So far, the plane wave orientations seem to be selected on a number theoretical basis only.
The relation to moiré patterns, however, becomes obvious when considering the symmetries in the distribution of the concyclic points (Eq.~\eqref{eq:concyclic}).

The units $\mathbb{Z}[\zeta_m ]^{\cross}$ are defined as lattice points with $N(\alpha)=\pm1$, where for periodic lattices the field norm can only be positive \cite{Washington_1997}.
Geometrically, these are the lattice points lying on the circle with radius $1$:
\begin{equation}
    \mathbb{Z}[\zeta_m]^{\cross} = \begin{cases} \{\pm 1, \pm i\} & \text{for }m=4 \\ \{\pm 1, \pm \zeta_{6}, \pm \zeta_{6}^2\} & \text{for }m=6\,. \end{cases}
\end{equation}
We can establish that the calculated intensity pattern using these points defines a certain ``base pattern.''
Also, the number of concyclic points $P(n)$ come in multiples of the number of units \cite{Cox_1989}.
Thus, using orientations $\arg[P(n)]$ is analogous to rotating multiple copies of the base pattern by \emph{twist} angles.
As a consequence, the integer lattice method is equivalently described in terms of moiré rotations.

For example, take the factorization $7=(3-\zeta_6)(3-\bar\zeta_6)$ over $\mathbb{Z}[\zeta_6]$, written in terms of its complex conjugate pair~\cite{Conway_1996}.
The set of concyclic points is therefore
\begin{equation}
    P(7) = U\cdot(3-\zeta_6)\cup U\cdot(3-\bar\zeta_6)\,,
\end{equation}
where $U=\mathbb{Z}[\zeta_6]^{\cross}$ is the set of the $6$ units, giving $12$ lattice points (Fig.~\ref{fig:mbi}d).
Setting $G=P(7)/\sqrt{7}$ when computing the pattern, results in a moiré superlattice with a triangular unit cell, as shown in Fig.~\ref{fig:mbi}e.

Turning the search for plane wave orientations into a factoring problem allows computation of high-order superlattices without the need for an exhaustive search for all possible circle radii.
Here, the order is considered to be the number of occurrences of the base pattern.
Let \mbox{$n=91$}, the factorization of this integer reveals that the moiré superpattern is determined by the factors $91=7\cdot13=(3-\zeta_6)(3-\bar\zeta_6)(4-\zeta_6)(4-\bar\zeta_6)$, such that:
\begin{align}
    P(91) = U & \cdot (3-\zeta_6)(4-\zeta_6) \cup U \cdot (3-\zeta_6)(4-\bar\zeta_6) \\ & \cup U \cdot (3-\bar\zeta_6)(4-\zeta_6) \cup U \cdot (3-\bar\zeta_6)(4-\bar\zeta_6) \nonumber\,,
\end{align}
corresponding to $24$ plane wave orientations forming the interference pattern.
The resulting coherent lattices retain the integer lattice symmetry but have varying periodicity, as is illustrated in Fig.~\ref{fig:size}a for $n=7$ and Fig.~\ref{fig:size}b for $n=91$.
More complex patterns can now be found systematically.
For example, $n=1729=7\cdot13\cdot19$ corresponds to $48$ lattice points (Fig.~\ref{fig:size}c).

Calculating the periodicity relies on a key result from moiré theory for plane wave superpatterns.
The final superposition of all plane waves contains the frequency components of each wave as well as the new frequencies obtained in each convolution.
Therefore, the dominant spatial features are determined by the components closest to the origin in the frequency domain~\cite{Amidror_2009}.
Identifying the integer lattice with the frequency domain, the pattern periodicity $d$ can be written in terms of lattice points $p_i\in P(n)$ as follows
\begin{equation}
    d = \left|\min \left\{ \frac{p_i - p_j}{\sqrt{n}} \right\}\right|^{-1}\,, \qquad i\neq j
\end{equation}
where the elements of the reciprocal space $p_i-p_j$ are normalized by $\sqrt{n}$ to have unit length.
This shows that the coherent lattice periodicity is proportional to the square root of the field norm $n$.

\begin{figure}[!h]
    \centering
    \includegraphics[width=0.9\linewidth,keepaspectratio]{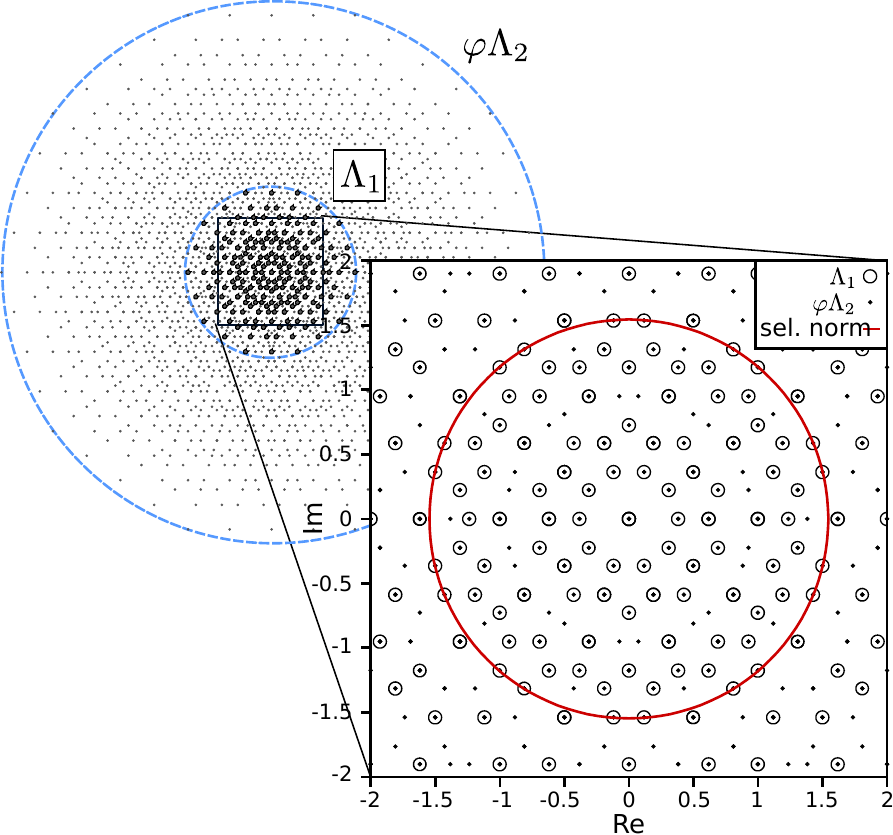}
    \caption{Illustration of the self-similarity property of quasilattices.
    The lattice points of the reduced decagonal quasilattice $\Lambda_1$ (circled dots) are a subset of the scaled quasilattice $\varphi\Lambda_2$ (dots), where $\varphi$ is the golden ratio. A selected circle (red) intersects with 20 lattice points.}
    \label{fig:self_similarity}
\end{figure}
\begin{figure*}[!ht]
    \centering
    \includegraphics[width=\linewidth,keepaspectratio]{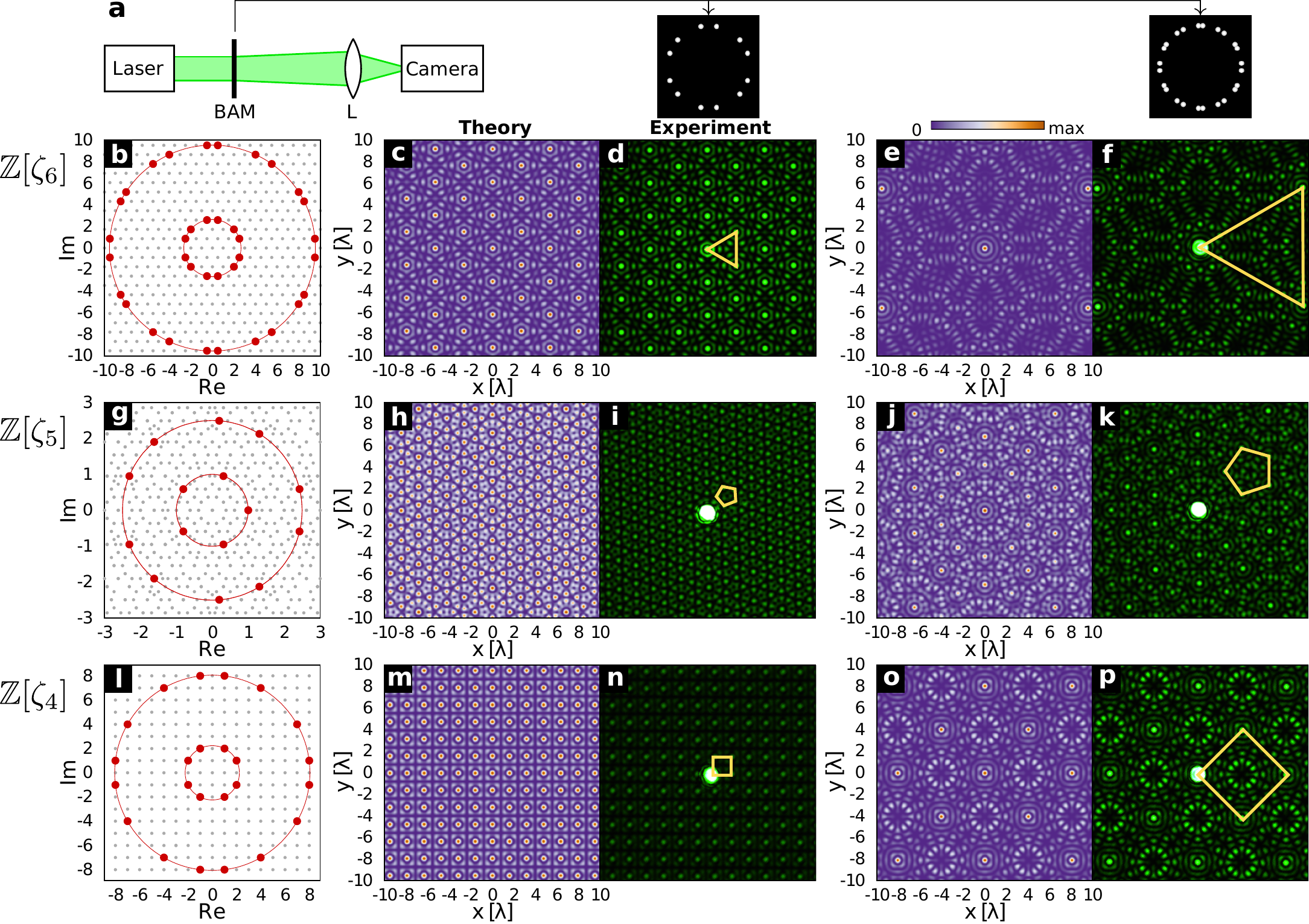}
    \caption{Experimental generation of optical coherent lattices for several key examples.
    (a) Setup: a laser beam passes through a binary amplitude mask (BAM) and the resulting diffraction pattern is focused on the camera using a lens (L).
    The BAM aperture positions are calculated using the integer lattice method, illustrated in the leftmost column.
    The values of the field norm $n=7$ and $n=91$ were selected for $\mathbb{Z}[\zeta_6]$ (b); $n=1$ and $n=4+\sqrt{5}$ for $\mathbb{Z}[\zeta_5]$ (g); and $n=5$ and $n=65$ for $\mathbb{Z}[\zeta_4]$ (l).
    Selected points, intensity plots, and experimental data are shown for square (b-f), fivefold quasiperiodic (g-k), and triangular patterns (l-p).
    For the fivefold-periodic case only half of the points were selected (g) since $\mathbb{Q}(\zeta_m)=\mathbb{Q}(\zeta_{2m})$.}
    \label{fig:interferograms}
\end{figure*}

\emph{Quasilattices} ---
Computing discrete lattice points becomes increasingly difficult for nonperiodic patterns.
In general, a lattice $\Lambda$ is a discrete subgroup generated by linearly independent elements.
That is, $\Lambda=\{\sum_ka_kl_k\mid a_k\in\mathbb{Z}\}\,,$ where $\{l_k\}$ is the set of generators \cite{Tkhang_1993}.
When considering quasiperiodic lattices, also called quasilattices, their aperiodic nature allows for infinitely many lattice points for a given norm.
This is also reflected by computing the units, as for example, for $\alpha\in\mathbb{Z}[\zeta_5]$, setting the norm $N(\alpha)=(a^2-5b^2)/4=\pm1$, gives infinitely many integer solutions $a, b$, complicating the integer lattice method for quasilattices.

A way to circumvent this behavior and be able to analyze a finite set of lattice points, is to limit the number of linear combinations of the generators by restricting the possible values of $a_k$.
We can do this by writing a reduced form of the quasilattice as follows:
\begin{equation}
    \label{eq:reduced_lattice} \Lambda_{n} = \left\{\sum_k a_k \zeta^k \,\middle\vert\, |a_k|\leq n, a_k \in \mathbb{Z} \right\}\,,
\end{equation}
where $\{\zeta\}$ are the generators of the cyclotomic field and $n$ can be chosen to be a positive integer, such that $\displaystyle\lim_{n\rightarrow\infty}\Lambda_{n} = \Lambda$, yielding the full quasilattice.

Moreover, quasilattices have the remarkable structural property of self-similarity, meaning that scaling the lattice points by a certain factor results in quasilattices that look identical \cite{Niizeki_1989}.
Let $\varphi$ be the scaling factor, from self-similarity it follows that for certain $n\in\mathbb{N}$ we have $\Lambda_n\subset\varphi\Lambda_{n+1}$.
Using this property and the fact that $\lim_{n\rightarrow\infty}\Lambda_{n}=\Lambda$, it is possible to construct arbitrary large finite integer quasilattices.
As an example, consider the ring $\mathbb{Z}[\zeta_{5}]$, where a good choice for $\varphi$ is the golden ratio $(1+\sqrt{5})/2$, which is the decagonal quasilattice self-similarity ratio~\cite{Niizeki_1989}.
The construction of this reduced quasilattice, illustrating the concept of self-similarity, is shown in Fig. \ref{fig:self_similarity}.

Returning to the problem of finding concyclic lattice points, such that these could be related to plane wave orientations, it should be noted that the field norm can have infinitely many corresponding points.
Therefore, lattice points must be grouped by their complex magnitude.
A set of concyclic points selected using this approach is shown in Fig.~\ref{fig:self_similarity}.
Incidentally, the field norm can have noninteger values for quasilattices.

\emph{Experimental results} ---
An experimental validation of the integer lattice method was obtained by generating optical coherent lattices as diffraction patterns emerging from a binary amplitude mask (Fig.~\ref{fig:interferograms}a).
Linearly polarized monochromatic light cones are derived from a common collimated laser beam ($\lambda=532\,\si{\nano\meter}$) by passing it through $10\,\si{\micro\meter}$ diameter circular apertures in an otherwise opaque mask, each aperture generating a diffraction cone (Fig.~\ref{fig:mbi}c).
The mask was prepared using UV lithography and lift-off on a $200\,\si{\nano\meter}$ sputtered chromium layer on a glass slide.
The placement of the apertures is determined by the concyclic points found from the integer lattice method.

To exemplify the integer lattice method, the impact of the choice of the field norm is demonstrated for triangular patterns (Fig.~\ref{fig:interferograms}b-f), quasiperiodic patterns with fivefold symmetry (Fig.~\ref{fig:interferograms}g-k), and square patterns (Fig.~\ref{fig:interferograms}l-p).
The measured interference patterns show good agreement with the calculated coherent lattices.
An increase in the primitive cell size is clearly obtained for increasing selected field norm with more prime factors.
Patterns like these can significantly impact cold atom physics and optical tweezer technology by offering access to nontrivial trapping potentials, as well as multifocal microscopy and imaging by allowing custom optical sectioning.

\emph{Conclusion} ---
We have presented a novel approach to design \mbox{(quasi-)periodic} interference patterns generated by coherent waves, with a focus on electromagnetic waves.
The formation of the patterns was shown to follow from prime number factorization, i.e. transforming wave components according to a selection of algebraic integers with equal field norm.
Furthermore, we have demonstrated that coherent lattices can be interpreted as moiré superlattices with periodicity bounded by the square root of the field norm.
The method was experimentally verified by forming key optical lattices as diffraction patterns in an optical setup.
We conclude that the integer lattice method is a powerful approach to design coherent lattices with great control over the spatial characteristics, such as pattern symmetry and spacing of the intensity maxima.

The discussion was restricted to two dimensions, but integer factorization and the moiré principle are equally valid for any dimension, including 3D lattices.
We postulate that an approach using geometrical algebra might suit a more general theoretical framework.
The proposed method was shown not only to be a robust way to design \mbox{(quasi-)periodic} patterns, but most importantly, opens the door to new ways of studying interference patterns based on algebraic number theory.
For example, a more thorough study of the field norm of integer quasilattices would be the next building block in the theoretical framework.

\begin{acknowledgments}
    We would like to thank Hamideh Jafarpoorchekab for providing help with sample fabrication.
    We acknowledge grant support from FWO Vlaanderen (No. 12ZR720N) to Q.D.
    This work is part of a project that has received funding from the European Research Council (ERC) under the European Union's Horizon 2020 research and innovation programme (Grant Agreement No. 805222).
\end{acknowledgments}

\bibliographystyle{apsrev4-2}
\end{document}